\documentclass[10pt,conference]{IEEEtran}
\IEEEoverridecommandlockouts

\usepackage{cite}
\usepackage{amsmath,amssymb,amsfonts}
\usepackage{algorithmic}
\usepackage{graphicx}
\usepackage{textcomp}
\usepackage{xcolor}
\usepackage{hyperref}
\def\BibTeX{{\rm B\kern-.05em{\sc i\kern-.025em b}\kern-.08em
    T\kern-.1667em\lower.7ex\hbox{E}\kern-.125emX}}
\begin{document}


\title{Combining Static Analysis Techniques for Program Comprehension Using Slicito\\
\thanks{This work has been supported by the Czech Science Foundation project no.~23-06506S and the project Cooperatio SCI-COMP.}
}

\author{\IEEEauthorblockN{Robert Husák}
\IEEEauthorblockA{\textit{Charles University}\\
\textit{Faculty of Mathematics and Physics} \\
Prague, Czech Republic \\
robert.husak@matfyz.cuni.cz}
\and
\IEEEauthorblockN{Jan Kofroň}
\IEEEauthorblockA{\textit{Charles University}\\
\textit{Faculty of Mathematics and Physics} \\
Prague, Czech Republic \\
jan.kofron@matfyz.cuni.cz}
\and
\IEEEauthorblockN{Filip Zavoral}
\IEEEauthorblockA{\textit{Charles University}\\
\textit{Faculty of Mathematics and Physics} \\
Prague, Czech Republic \\
filip.zavoral@matfyz.cuni.cz}
}

\maketitle

\begin{abstract}
While program comprehension tools often use static program analysis techniques to obtain useful information, they usually work only with sufficiently scalable techniques with limited precision.
A possible improvement of this approach is to let the developer interactively reduce the scope of the code being analyzed and then apply a more precise analysis technique to the reduced scope.
This paper presents a new version of the tool \textsc{Slicito} that allows developers to perform this kind of exploration on C\# code in Visual Studio.
A common usage of \textsc{Slicito} is to use interprocedural data-flow analysis to identify the parts of the code most relevant for the given task and then apply symbolic execution to reason about the precise behavior of these parts.
Inspired by Moldable Development, \textsc{Slicito} provides a set of program analysis and visualization building blocks that can be used to create specialized program comprehension tools directly in Visual Studio.
We demonstrate the full scope of features on a real industrial example both in the text and in the following video: \url{https://www.slicito.com/icpc2025video.mp4}
\end{abstract}


\section{Introduction}

Static program analysis comprises techniques that reason about program behavior by inspecting its source code.
They are often used in program comprehension tools~\cite{DeeringAtlas,CrichtonFlowistry,LaTozaReacher} to obtain additional information for developers.
These techniques differ by their precision and scalability; precise techniques tend not to be very scalable, and vice versa.

Program comprehension tools usually employ reasonably scalable techniques to ensure that their responses to developers' questions are reliable.
The downside of this approach is that the precision of the selected technique limits the tool's precision.
The approach presented in this paper allows developers to combine several techniques of varying precision.
At first, the developer uses a scalable technique to restrict the scope to the relevant code.
Then, they obtain detailed information using a precise technique.

This paper describes how we implemented this approach into the latest version of a tool called \textsc{Slicito}\footnote{\label{foot:url}\url{https://www.slicito.com}, \url{https://github.com/roberthusak/slicito}}.
It evolved from a set of libraries to analyze and visualize C\# source code using computational notebooks in Visual Studio Code~\cite{SlicitoIcpc2023} to a Visual Studio extension that provides data-flow analysis and symbolic execution.

Section~\ref{sec:features} compares the current version of \textsc{Slicito} with the original one and lists the most important features.
Section~\ref{sec:example} illustrates the tool's usage on a real-life example.
Section~\ref{sec:architecture} describes the internals of \textsc{Slicito} and the fundamental decisions about its architecture.

\begin{figure*}[h]
\centering
\includegraphics[width=\textwidth]{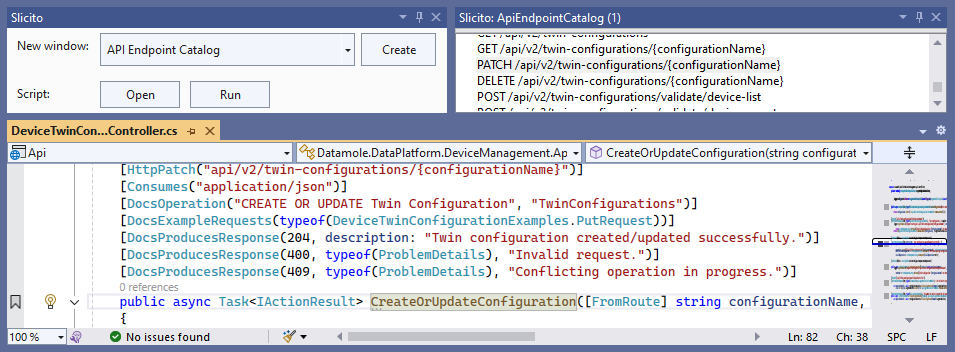}
\caption{Integration of \textsc{Slicito} into Visual Studio.}
\label{fig:example:1}
\end{figure*}

\section{Features}
\label{sec:features}

\textsc{Slicito} was initially created as a set of libraries for computational notebooks, documents consisting of text annotations, code cells, and the results of code execution~\cite{SlicitoIcpc2023}.
By writing code snippets into a notebook and visualizing their results, a developer could explore the structure of the project and the relations between namespaces, classes, and other entities.
While the basic idea of short code snippets and information visualization still holds, the version of \textsc{Slicito} presented in this paper was overhauled entirely to address numerous issues from the original implementation.

Since Visual Studio is the most popular integrated development environment (IDE) for C\#, \textsc{Slicito} has always used it to navigate the developer from a generated schema to a specific location in the code.
However, the developer had to write code snippets in a separate Visual Studio Code window because Visual Studio does not support computational notebooks.
To simplify this process, the current version of \textsc{Slicito} no longer targets computational notebooks.
Instead, it runs directly in Visual Studio and executes code snippets written in the scripting dialect of C\#.
The resulting visualizations are displayed in integrated tool windows.

As \textsc{Slicito} enables developers to gradually restrict the scope of the code they are reasoning about, the underlying abstraction model was extended to handle this process more intuitively using the notion of \textit{slices} that adhere to specific constraints.
For example, when the developer includes a particular method in a slice, its containing class and namespaces are also automatically included in the slice.
See Section~\ref{sec:architecture} for more details.

The restructuring was also an opportunity to split the implementation into multiple loosely coupled parts with clear interfaces.
For example, working directly with the types provided by the .NET Compiler Platform (``Roslyn'') was cumbersome.
Therefore, they are now converted to a simplified representation used in intraprocedural data-flow analysis and symbolic execution.

Although developers can run these analysis techniques directly, it is expected to be too complicated for most of them.
Therefore, \textsc{Slicito} provides several high-level analysis and visualization techniques that should be more understandable by the general developer audience.
\textit{Call graph analysis}, as its name suggests, constructs a call graph from a set of entry methods.
The developer can display the call graph using \textit{call graph explorer}, which can optionally emphasize methods that are data-dependent on a selected parameter from an entry method.
The visualizer obtains interprocedural data dependency by performing data-flow analysis on individual methods and combining their results.
The power of symbolic execution is provided by \textit{reachability analysis}, which allows the developer to specify the conditions for the parameters and the return value of a method and obtain the values of inputs under which these conditions are satisfiable.

Two visualizers are not dependent on static program analysis.
\textit{Structure browser} enables the developer to interactively explore a slice's contents, similarly to the interface for exploring \textit{contexts} in the original version.
\textit{API endpoint catalog} is an example of a tool created for a specific domain.
As its name suggests, it produces a list of endpoints published by an ASP.NET Core application and provides links to the methods that handle them.

Given that \textsc{Slicito} is currently a research prototype, only a limited subset of C\# constructs and types is supported.
The project documentation\textsuperscript{\ref{foot:url}} contains more detailed information.

\section{Example Usage}
\label{sec:example}

We demonstrate how \textsc{Slicito} can be used in practice on an actual task from industry.
The system under question is the Spotflow Internet of Things (IoT) Platform\footnote{\url{https://spotflow.io}}, responsible for the monitoring, management, and configuration of tens of thousands of IoT devices.
The platform runs on Microsoft Azure and consists of several microservices that can be managed via an HTTP application programming interface (API).
The task was to inspect all the API endpoints responsible for creating platform assets and verify that the names of these assets undergo appropriate validation.
There were 10 asset types to check; we show only the work associated with the asset of the type \textit{configuration}, whose name must be 1 to 64 characters long, contain only the character ranges \texttt{0-9}, \texttt{a-z}, and dashes, and start and end with an alphanumeric character.

After the developer opens the solution of the relevant microservice in Visual Studio, the first step is to find the code responsible for responding to the configuration creation endpoint.
Normally, this would mean either exploring the code manually or full-text searching the fragments of the endpoint URL.
Since this task is common for ASP.NET Core developers, \textsc{Slicito} provides a built-in tool window that lists all the endpoints and navigates the developer to the corresponding code when the endpoint is selected.

Fig.~\ref{fig:example:1} shows what the interaction in Visual Studio looks like.
When the developer clicks the button \textit{Create}, the selected type of tool window is created, such as the window \textit{Slicito: ApiEndpointCatalog} in our case.
Clicking the appropriate API endpoint navigates the developer to the corresponding handler method in the code.

\begin{figure*}
\centering
\includegraphics[width=\textwidth]{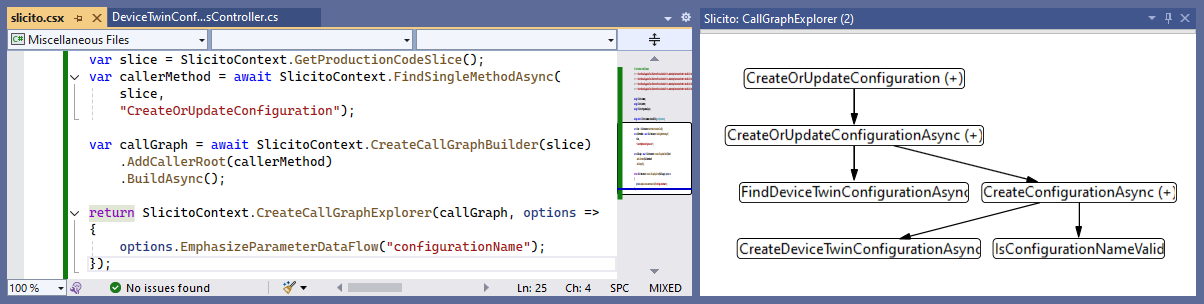}
\caption{Creating a customized interactive tool window with \textsc{Slicito}.}
\label{fig:example:2}
\end{figure*}

The next step is to find the location where the configuration name format is checked.
Instead of reading the code, the developer can use \textsc{Slicito} to create a simple custom tool for this purpose.
Clicking the button \textit{Open} in the top window from Fig.~\ref{fig:example:1} creates a new C\# script file.
The button \textit{Run} runs the script and opens a new tool window based on the return value.
Fig.~\ref{fig:example:2} shows a script that creates a call graph with the endpoint handler as its only root and displays it in the \textit{CallGraphExplorer} tool window.
The explorer is instructed to expand the methods where the configuration name is used.
The developer can now directly see that there is a special method \texttt{IsConfigurationNameValid} that verifies the format of the configuration name using a regular expression before creating it using \texttt{CreateDeviceTwinConfigurationAsync}.

\begin{figure}
\centering
\includegraphics[width=\columnwidth]{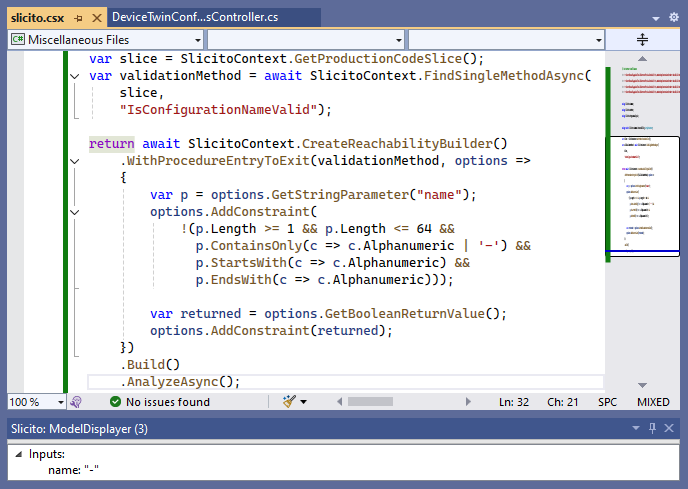}
\caption{Using symbolic execution in \textsc{Slicito}.}
\label{fig:example:3}
\end{figure}

Now, it remains to verify that the regular expression performs the validation correctly, i.e., that no name not satisfying the conditions is allowed in the code.
Because the scope of the code under consideration has now been reduced to a great extent, it is possible to use symbolic execution for this purpose.
Fig.~\ref{fig:example:3} captures how the developer extended the previously mentioned C\# file to perform this check and display the results in a new tool window.
\texttt{ReachabilityAnalysis} uses symbolic execution to precisely assess whether the input value constrained by the given conditions prevents the reachability of the given position in the program.
If any such input values exist, they are displayed in the tool window.
In our example, the configuration name consisting of a single dash is unintentionally allowed.

Eventually, the developer fixes the regular expression and runs the last check again to ensure the name validation works correctly.
Since checks must be performed for the remaining nine asset types, it is now beneficial to reuse the existing script with slight modifications.

\section{Architecture}
\label{sec:architecture}

\begin{figure}
\centering
\includegraphics[width=0.9\columnwidth]{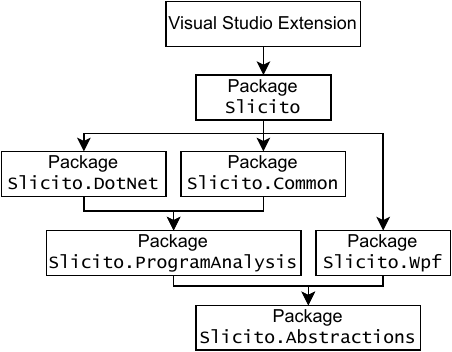}
\caption{Dependencies between the components of \textsc{Slicito}.}
\label{fig:architecture}
\end{figure}

Although we present \textsc{Slicito} mainly as a Visual Studio extension for C\#, most of its features are implemented in a set of NuGet packages.
As a result, developers can use these features, for example, in standalone console applications and libraries.
Because each package addresses a different concern, the set of supported IDEs, programming language, and static program analysis techniques can be extended with reasonable effort.
Fig.~\ref{fig:architecture} shows the dependencies between the components \textsc{Slicito} consists of.

The NuGet package \texttt{Slicito.Abstractions} contains the fundamental data types and interfaces all other libraries use.
Since an important feature is the ability to restrict the program's scope to the parts relevant to the particular task, we provide an abstraction to represent the program structure.
The abstraction is based on graph theory because many static analysis techniques already use graphs to represent programs~\cite{DragonBook,FerrantePdgOptimization,HorwitzInterproceduralSlicing,Cpgs}.
In particular, we use a directed multigraph with loops whose nodes and edges can be annotated with additional attributes.
We call the nodes of this multigraph \textit{elements} and its edges \textit{links}; these two kinds of entities together are called \textit{facts}.
Multigraph instances are called \textit{slices}.
While the name references program slicing~\cite{WeiserProgramSlicing}, a slice must not necessarily be an output of a program slicing algorithm.
For instance, a helpful slice is the subset of the projects that will be deployed to production, that is, all the code except tests and examples.

Other important abstractions in this package are \textit{controllers} and \textit{models} that serve as the base for the development of custom tools with a graphical user interface (GUI).
The implementation of these tools follows the model-view-controller (MVC) pattern~\cite{fowler2003patterns}.
The idea is that the tool implementors create custom controllers that provide and receive information using a set of pre-defined models.
An example of a model is a nested list of items, each with a specified action sent to the controller when the developer clicks it.
Since individual IDEs use different libraries and often even different programming languages for their GUIs, \textit{views} are implemented in separate packages.
Currently, the only implementation of views is in the package \texttt{Slicito.Wpf}.
This package implements the views in Windows Presentation Foundation (WPF) used in Visual Studio.

All the remaining NuGet packages are independent of the visual representation.
\texttt{Slicito.ProgramAnalysis} contains the implementation of the supported static program analysis techniques: data-flow analysis~\cite{KildallProgramOptimization,DragonBook} and symbolic execution~\cite{KingSE,BaldoniSurvey}.
These techniques are performed on an intermediate representation of program semantics that is not tied to a particular programming language.
The representation is based on control flow graphs (CFGs) and expression trees.

\texttt{Slicito.DotNet} implements the abstractions provided by its underlying packages to extract the structure and semantics of C\# programs.
The extraction of facts and the construction of CFGs for static analysis are performed lazily on request from other components.
The package also manages the mapping between the extracted methods and their corresponding CFGs.

\texttt{Slicito.Common} comprises the implementation of three important features.
First, it implements the remaining abstractions from \texttt{Slicito.Abstractions} that are not specific to any programming language.
Second, it provides a collection of general-purpose controllers, such as call graph explorer.
Third, it contains an engine that dynamically loads the controllers' implementations from executed files written in the scripting dialect of C\#.

The Visual Studio extension references all the packages via the package \texttt{Slicito} that serves as their aggregator.
The extension ensures that the dependencies of all used classes are fulfilled and integrates user interaction via the MVC pattern into Visual Studio's GUI.
Because symbolic execution needs a satisfiability modulo theories (SMT) solver, the extension contains the binary executable file of Z3~\cite{Z3}.
It also uses the engine for dynamic loading of dependencies to allow the developer to create new controllers using short C\# code snippets.

\section{Related Work}

The ability of developers to build their custom program comprehension tools during the software development process is the fundamental principle of Moldable Development~\cite{MoldableDevelopment}.
Our approach differs from theirs in that we focus on static program analysis and use customized abstractions for this purpose.
Although the authors of Moldable Development recommend using tools for analyzing source code as well, their primary focus is on the dynamic interaction with the developer.
The IDE \textsc{Glamorous Toolkit} provided for this purpose is implemented in a dynamically typed language called Pharo, which would be hard to use by most C\# developers.

Multiple program comprehension tools use static program analysis techniques combined with visualization and configurable queries.
The \textsc{Solid*} toolset from Reniers et al.~\cite{SolidToolset} extracts the facts from the code, processes them, and displays them in a selected graph according to a specific query.
\textsc{NDepend}\footnote{\url{https://www.ndepend.com}} is a commercial tool for the analysis of code metrics and dependencies in .NET projects that provides a custom query language.
\textsc{Atlas} from Ensoft Corp.~\cite{DeeringAtlas} can extract control and data flow from Java code and visualize it interactively.
\textsc{Reacher} from LaToza and Myers~\cite{LaTozaReacher} uses a custom static analysis technique called Fast Feasible Path Analysis (FFPA)~\cite{LaTozaDissertation} to improve the precision of the responses to reachability questions in Java code.
Crichton's \textsc{Flowistry}~\cite{CrichtonFlowistry} is an extension for Rust that performs program slicing to highlight the code relevant to the currently selected statement in a function.

\textsc{Slicito} differs from these tools by gradually increasing the precision of the analysis as the scope decreases until it is possible to use techniques such as symbolic execution.

\section{Conclusion}

This paper describes the features and architecture of the Visual Studio extension \textsc{Slicito} and demonstrates how developers can use it to obtain detailed information about selected parts of their programs.
\textsc{Slicito} supports interprocedural data-flow analysis and symbolic execution of C\# programs.
Developers can use these techniques using an intuitive API to create custom interactive tool windows specific to their tasks.

Since \textsc{Slicito} is currently in the research prototype phase, the next step is to extend the set of supported C\# constructs and .NET libraries to make it usable for a larger scope of tasks.
Then, we want to evaluate its usefulness by using it within a real software development team and measuring its influence on the development process.
Further plans include extending the supported programming languages and static analysis techniques.
Another interesting direction is to evaluate if an interface based on LLMs could help developers use the features of \textsc{Slicito} more easily.
If this solution indeed works, it would combine the benefits of both parts: the convenience of an LLM interface and the reliability of static program analysis.

\bibliographystyle{IEEEtran}
\bibliography{IEEEabrv,paper}

\end{document}